\documentclass[fleqn,usenatbib]{mnras}

\usepackage[T1]{fontenc}

\DeclareRobustCommand{\VAN}[3]{#2}
\let\VANthebibliography\thebibliography
\def\thebibliography{\DeclareRobustCommand{\VAN}[3]{##3}\VANthebibliography}

\usepackage{graphicx}	
\usepackage{amsmath}	
\usepackage{amssymb}	

\title[An Explicit Solution for Spherical Collapse]{A Uniform Spherical Goat (Problem): Explicit Solution for Homologous Collapse's Radial Evolution in Time}

\author[Z. Slepian \& O. Philcox]{
Zachary Slepian$^{1,2}$\thanks{E-mail: \href{mailto:zslepian@ufl.edu}{zslepian@ufl.edu} (ZS)}
and Oliver H.\,E. Philcox$^{3,4}$\thanks{E-mail: \href{mailto:ohep2@cantab.ac.uk}{ohep2@cantab.ac.uk} (OP)}
\\
$^{1}$Department of Astronomy, University of Florida, 211 Bryant Space Science Center, Gainesville, FL 32611, USA\\
$^{2}$Physics Division, Lawrence Berkeley National Laboratory, 1 Cyclotron Road, Berkeley, CA 94709, USA\\
$^{3}$Department of Astrophysical Sciences, Princeton University, Princeton, NJ 08540, USA\\
$^{4}$School of Natural Sciences, Institute for Advanced Study, 1 Einstein Drive, Princeton, NJ 08540, USA}

\pubyear{2021}

\usepackage{color}
\definecolor{darkgreen}{RGB}{0,120,0}

\newcommand{\oliver}[1]{\textcolor{red}{(\textbf{Oliver}: #1)}}
\newcommand{\new}[1]{#1}

\usepackage{newtxtext,newtxmath}
\begin{document}
\label{firstpage}
\pagerange{\pageref{firstpage}--\pageref{lastpage}}
\maketitle
\begin{abstract}
The homologous collapse from rest of a uniform density sphere under its self gravity is a well-known toy model for the formation dynamics of astronomical objects ranging from stars to galaxies. Equally well-known is that the evolution of the radius with time cannot be explicitly obtained because of the transcendental nature of the differential equation solution. Rather, both radius and time are written parametrically in terms of the development angle $\theta$. We here present an explicit integral solution for radius as a function of time, exploiting methods from complex analysis recently applied to the mathematically-similar ``geometric goat problem.'' Our solution can be efficiently evaluated using a Fast Fourier Transform and allows for arbitrary sampling in time, with a simple \textsc{python} implementation that is $\sim$$100\times$ faster than using numerical root-finding to achieve arbitrary sampling. Our explicit solution is advantageous relative to the usual approach of first generating a uniform grid in $\theta$, since this latter results in a non-uniform radial or time sampling, less useful for applications such as generation of sub-grid physics models.
\end{abstract}


\section{Introduction}
\label{sec:intro}
Spherical collapse is ubiquitous in astronomy and has been used to model the formation of stars up to the formation of galaxy halos. Going back many years (e.g. \citealt{LMS65,Tomita, Gunn}), the model of a uniform density sphere collapsing homologously (no shell crosses another shell) from rest under its own gravity has been the simplest instantiation of this scenario. The governing equations (see e.g. \citealt{LMS65}) are
\begin{align}
&\frac{d^2 r}{dt} = -\frac{GM_r}{r^2} = -\frac{4\pi G \rho_0} {3}\frac{r_0^3}{r^2},\nonumber\\
&M_r \equiv \frac{4\pi}{3} r_0^3 \rho_0,
\end{align}
where $r$ is the radius of the sphere, $G$ is Newton's constant, $M_r$ is the mass internal to radius $r$, and $\rho_0$ is the (uniform) initial density when the sphere has its initial radius, $r_0$. The second equality in the first line above comes from inserting the form for $M_r$ given in the second line. These have the cycloidal
parametric solution
\begin{align}
    &r(\theta) = r_0 \cos^2 \theta,\nonumber\\
    &\theta + \frac{1}{2} \sin 2\theta = \frac{\pi}{2} \frac{t}{t_{\rm ff}},\qquad t_{\rm ff} \equiv \sqrt{\frac{3\pi}{32 G \rho_0}},
    \label{eqn:theta_eqn}
\end{align}
where $t_{\rm ff}$ is the free-fall time. $\theta=0$ corresponds to the initial conditions of radius $r_0$ and zero velocity, and at $\theta = \pi/2$, the sphere has collapsed to zero radius.
Since the equation for $\theta$ is transcendental, one cannot explicitly obtain $\theta$ as a function of $t$ and thence $r(t)$. Here, we show how using techniques from complex analysis recently developed to solve the ``geometric goat problem'' (which we will momentarily describe), an explicit integral solution for $r(t)$ can be found.

The geometric goat 
problem is as follows. Suppose a goat is placed inside a circular (2-D) enclosure of radius $R$, tethered to a fixed point on the circumference by a rope of length $r$. How long must the rope be to permit the goat to graze on exactly half the area of the enclosure?

Writing down the appropriate integral expressions for the enclosed area as a function of $r$ and $R$, one obtains a transcendental equation. Following a number of (non-trivial) manipulations, this equation can be written as
\begin{align}
    \sin \beta - \beta \cos \beta = \frac{\pi}{2}
\end{align}
\citep{Ullisch}. We observe that this equation is somewhat similar to our equation (\ref{eqn:theta_eqn}) for $\theta$ if one treats $t$ as a constant and $\theta$ as analogous to $\beta$. If one is able 
to solve an equation of the type above, it is worth considering whether 
the same method may be used to solve equation \eqref{eqn:theta_eqn} for $\theta(t)$. This indeed turns out to be so. 

\section{Solution}\label{sec:soln}
We follow the approach \new{described in} \citet{Ullisch} to obtain our solution. 

First, we write our $\theta$ equation \eqref{eqn:theta_eqn} in terms of an \new{entire} function $f(z)$ defined on the complex plane, 
\begin{align}
    f(z) &\equiv z + \frac{1}{2} \sin 2z - \frac{\pi}{2} \frac{t}{t_{\rm ff}}.
    \label{eqn:f0_def}
\end{align}
\new{Here, we require $f(z_0) = 0$, where $z_0$ will give our desired solution $\theta(t)$. By symmetry, this has $\mathrm{Im}(z)=0$.}
For real $z$, and at fixed $t>0$, $f(z)$ is monotonically increasing on this interval and has exactly one zero (\textit{i.e.} one solution for $\theta(t)$). 
\new{This zero may be shown to be simple}:
\begin{align}
    \lim_{z \to z_0} \frac{z + (1/2) \sin 2z - (\pi/2)(t/t_{\rm ff})}{z - z_0} = 1 + \cos 2 z_0 \neq 0
\end{align}
\new{given the bounds on $z_0$}.

{\bf Theorem 1} of \citet{Ullisch} \new{(see also \citealt{Jackson16,Jackson17,Luck15})} states that, on a simply-connected open subset \new{$U$} of the complex plane, for every simple zero \new{$z_0\in U$} of a non-zero analytic function $f(z)$, there exists a curve $C$ such that
\begin{align}
    z_0 = \frac{\oint_C z\;dz/f(z)}{\oint_C dz/f(z)}.
    \label{eqn:thm}
\end{align}
\new{Indeed, this is true for \textit{any} Jordan curve $C$ (\textit{i.e.} one which is continuous and does not self-intersect), enclosing $z_0$ such that $z_0$ is the only zero of $f(z)$ on $C$ and its interior.}
\new{To apply this method to the spherical collapse scenario}, we must must thus find \new{a valid} curve $C$ by which to evaluate the result (\ref{eqn:thm}). 

\new{Motivated by the boundary conditions on $\theta$ and the discussion in \citet{Ullisch}, we first consider the (simply-connected) rectangular region $R = (0,\pi/2)\times i(-M,M)$ in the complex plane for arbitrary $M>0$. Via the argument principle, the number of zeros minus the number of poles contained within $R$ is given by
\begin{align}
    \frac{1}{2\pi i}\oint_{\partial R}\frac{f'(z)}{f(z)}dz = \frac{1}{2\pi}\times \Delta_{\partial R}[\arg f(z)]
\end{align}
where $\partial R$ is the (non-self-intersecting) boundary of $R$ (traversed counter-clockwise), and $\Delta_{\partial R}[\arg f(z)]$ represents the total change in the argument of $f(z)$ as one traverses $\partial R$.\footnote{\new{This is easily proven by noting that $\log f(z)$ is the antiderivative of $f'(z)/f(z)$ and using the relation between the complex logarithm of a function and the function argument.}} Given that $f(z)$ contains no poles in $R$, this simply counts the number of zeros within $R$.}

\new{Denoting $z = x+iy$, $f(z)$ has the limiting forms
\begin{align}
    f\left(0+iy\right) &= -\frac{\pi}{2}\frac{t}{t_{\rm ff}}+i\left(y+\frac{1}{2}\sinh 2y\right)\nonumber\\
    f\left(\frac{\pi}{2}+iy\right) &= -\frac{\pi}{2}\left(1-\frac{t}{t_{\rm ff}}\right)+i\left(y-\frac{1}{2}\sinh 2y\right)\nonumber\\
    f\left(x+iM\right) &\approx x-\frac{\pi}{2}\frac{t}{t_{\rm ff}}+\frac{i}{4}e^{-2ix}e^{2M}\nonumber\\
    f\left(x-iM\right) &\approx x-\frac{\pi}{2}\frac{t}{t_{\rm ff}}-\frac{i}{4}e^{2ix}e^{2M},
\end{align}
where the third and fourth equations are exact in the limit $M\rightarrow\infty$. Let us consider the change in $\arg f(z)$ along each of the four sides of $\partial R$ in turn (assuming $M\gg 0$).
\begin{enumerate}
    \item $(0+iM) \rightarrow (0-iM)$. $\mathrm{Re}[f(z)]$ takes the constant (negative) value $-(\pi/2)(t/t_{\rm ff})$, whilst $\mathrm{Im}[f(z)]$ decreases monotonically from $(1/4)\exp{2M}$ to $-(1/4)\exp{2M}$. Thus $\Delta[\arg f(z)] = +\pi$.\vskip 4 pt
    \item $(0-iM) \rightarrow (\pi/2-iM)$. For large $M$, $f(z)\approx (1/4)\exp\left[2M+2ix+3i\pi/2\right]$, thus $\Delta[\arg f(z)] = +\pi$ as $x$ increases from $0$ to $\pi/2$.\vskip 4 pt
    \item $(\pi/2-iM)\rightarrow(\pi/2+iM)$. $\mathrm{Re}[f(z)]$ takes the constant (positive) value $(\pi/2)(1-t/t_\mathrm{ff})$, whilst $\mathrm{Im}[f(z)]$ increases monotonically from $(1/4)e^{2M}$ to $(1/4)e^{-2M}$. Thus $\Delta[\arg f(z)] = -\pi$.\vskip 4 pt
    \item $(\pi/2+iM)\rightarrow(0+iM)$. For large $M$, $f(z)\approx (1/4)\exp\left[2M-2ix+i\pi/2\right]$, thus $\Delta[\arg f(z)] = +\pi$ as $x$ decreases from $\pi/2$ to $0$.\vskip 4 pt
\end{enumerate}
Summing the regimes, we find $\Delta_{\partial R}[\arg f(z)]=2\pi$, indicating that $R$ contains exactly one zero. Since the point $z_0\in R$, this must be the only zero in the region. Since $M$ is arbitrary, we can thus write $f(z)\neq 0$ for all $z\in U\backslash\{z_0\}$,\footnote{\new{$U\backslash\{z_0\}$ indicates the set $U$ excluding the point $z_0$.}} where $U = \{z:\mathrm{Re}(z)\in(0,\pi/2)$.}

\begin{figure}
    \centering
    \includegraphics[width=0.94\linewidth]{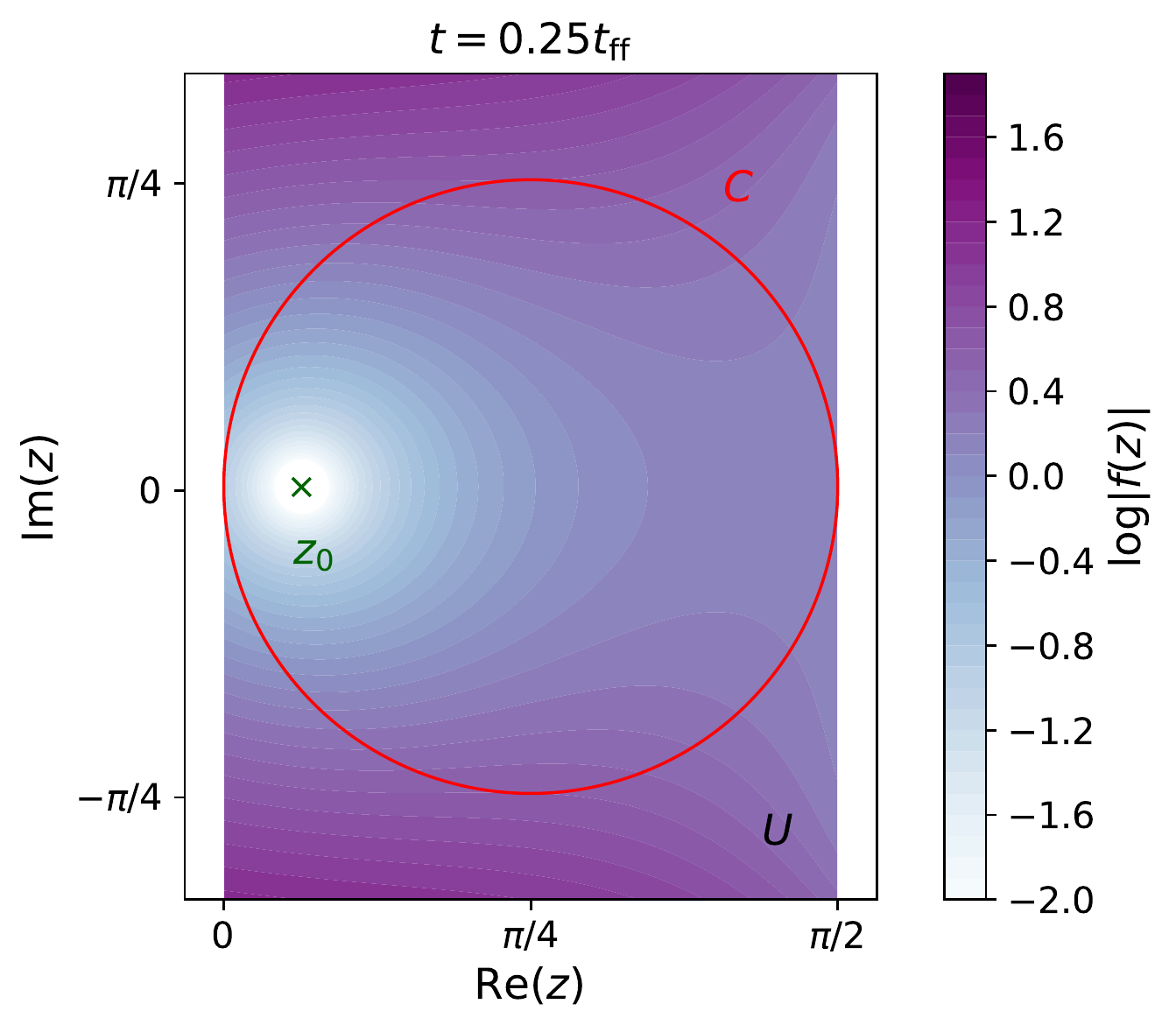}
    \caption{Plot of $f(z)$ (equation \ref{eqn:f0_def}) in the complex plane, from which the integral solution \eqref{eqn: integral-soln} is obtained. The colorbar shows the value of $|f(z)|$ with the green $\times$ indicating $z_0$ such that $f(z_0) = 0$. The colored area indicates the open subset $U$ upon which the theorem of \citet{Ullisch} is valid (denoted by the black $U$ at lower right), and we evaluate the contour integrals on the closed curve $C$ (in red). As proven in the text, $z_0$ is the sole zero of $f(z)$ in $U$. Here, we show the behavior for $t = 0.25t_\mathrm{ff}$ and $\epsilon = 10^{-4}$ (used to define $C$), but all choices are qualitatively similar.}
    \label{fig: contours}
\end{figure}

\new{Coupled with the theorem of \citet{Ullisch}, we see that any Jordan curve $C\in U$ enclosing $z_0$ can be used to evaluate equation \eqref{eqn:thm}. Here, we set $C$ equal to a circle with radius $\pi/4-\epsilon$ at center $(\pi/4,0)$ where $\epsilon>0$ is small. 
This is contained within $U$ and, for sufficiently small $\epsilon$, encloses $z_0$, thus the above conditions apply. A representative plot of $f(z)$, alongside the region $U$ and the contour $C$ is shown in Figure \ref{fig: contours}.}

We hence \new{obtain the integral solution}
\begin{align}
    &z_0(t) = \frac{\oint_{C} z\;dz/\left[z \new{+} (1/2) \sin 2 z - (\pi/2)(t/t_{\rm ff})\right]}{\oint_{C} dz/\left[z \new{+} (1/2) \sin 2 z - (\pi/2)(t/t_{\rm ff})\right]},\nonumber\\
    &r(t) = r_0 \cos^2 \left(\frac{\oint_{C} z\;dz/\left[z \new{+} (1/2) \sin 2 z - (\pi/2)(t/t_{\rm ff})\right]}{\oint_{C} dz/\left[z \new{+} (1/2) \sin 2 z - (\pi/2)(t/t_{\rm ff})\right]}\right),\nonumber\\
    & C =\{z: |z - \pi/4| = \pi/4 - \epsilon\}.\label{eqn: integral-soln}
\end{align}

\section{Evaluation using Fast Fourier Transforms}
\new{Following \citet{Ullisch}, we consider how to evaluate equation \eqref{eqn: integral-soln} using Fast Fourier Transforms (FFTs). First, we parametrize the contour $C$ by the function $\gamma(x) = \pi/4+(\pi/4-\epsilon)e^{2\pi i x}$ where $x\in[0,1]$ and we identify $\gamma(0)=\gamma(1)$. With this choice, the integral solution for $z_0$ becomes
\begin{align}
    z_0(t) &= \frac{\pi}{4}+\left(\frac{\pi}{4}-\epsilon\right)\frac{\int_{0}^1 dx\;e^{4\pi ix}\,g(x;t)}{\int_0^1 dx\;e^{2\pi ix}g(x;t)}\nonumber\\
    &=\frac{\pi}{4}+\left(\frac{\pi}{4}-\epsilon\right)\frac{c_{-2}(t)}{c_{-1}(t)},
    \end{align}
where $g(x;t)\equiv 1/f\left(\pi/4+(\pi/4-\epsilon)e^{2\pi i x}; t\right)$, making the $t$-dependence explicit, and we define the Fourier coefficients 
\begin{align}\label{eqn:ck-def}
    c_{k}(t) \equiv \int_0^1 dx\,g(x;t)e^{-2\pi i k x}
\end{align}
for integer $k$. The solution for $r(t)$ is thus
\begin{align}
    r(t) = r_0\cos^2\left(\frac{\pi}{4}+\left(\frac{\pi}{4}-\epsilon\right)\frac{c_{-2}(t)}{c_{-1}(t)}\right)
\end{align}
which can be computed to arbitrary precision for a given $t$ by estimating $c_{k}(t)$ using FFTs.\footnote{Alternatively, we may numerically integrate \eqref{eqn:ck-def} directly to compute only the $c_{-2}$ and $c_{-1}$ coefficients. In practice, this is slightly more efficient than using FFTs.} We recall that $\epsilon$ enters the radius of the contour $C$ (see Figure \ref{fig: contours}) used for the integration; this radius is $\pi/4 - \epsilon$.}

\begin{figure}
    \centering
    \includegraphics[width=80mm,scale=0.67]{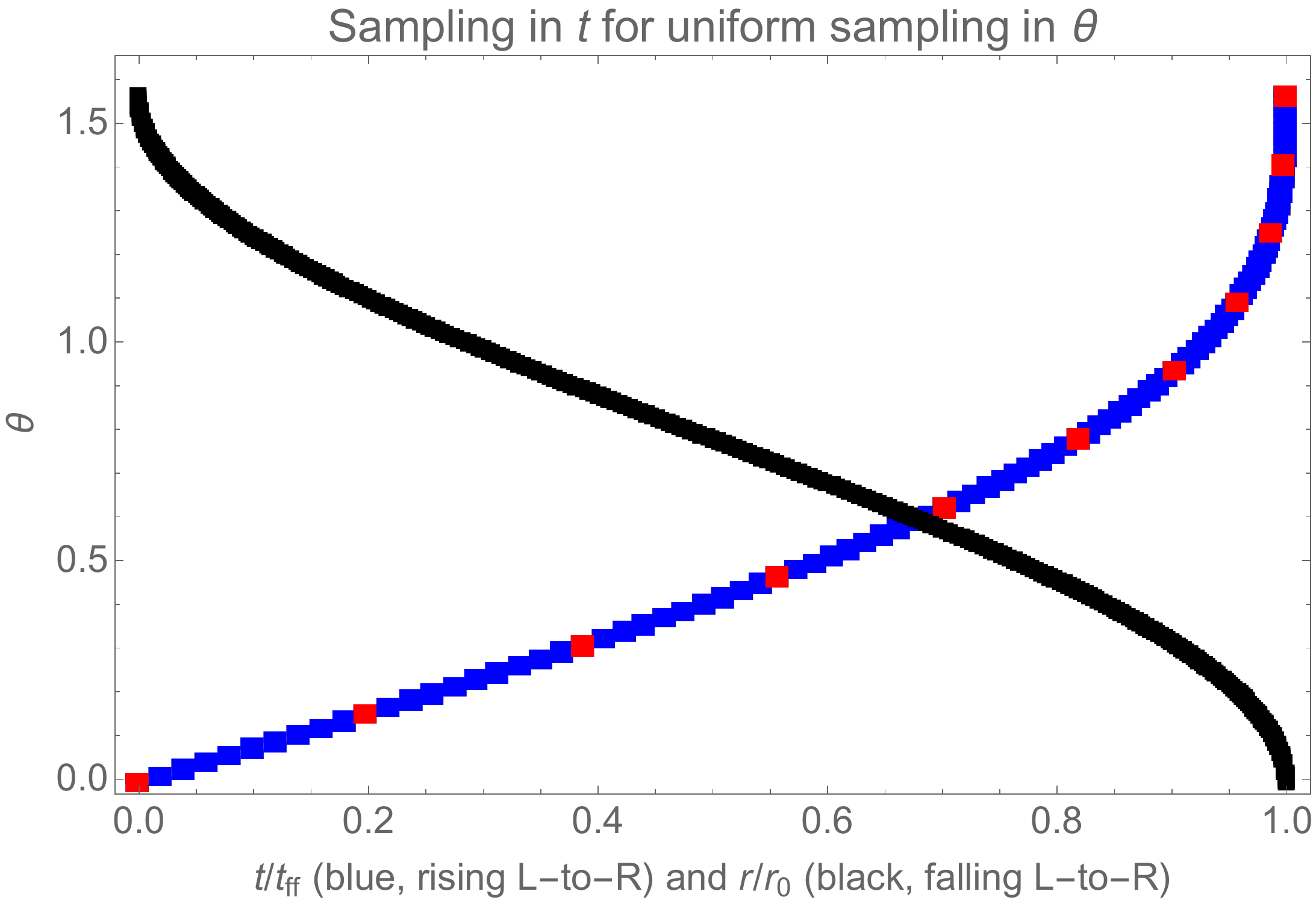}
    \caption{We show in blue (100 points) and red (10 points) the sampling in $t$ (horizontal axis) implied by uniform sampling in $\theta$ (vertical axis) of equation \eqref{eqn:theta_eqn}. This illustrates that the na\"ive numerical approach of taking a \new{uniformly-sampled} 1-D array in $\theta$ to evaluate the cycloid trajectory $r$ (black) does not necessarily provide good control of the sampling in $t$ for which one might wish. Here, we see that as $r/r_0 \to 0$ (moving towards the left on the horizontal axis) the uniform sampling in $\theta$ gives less and less good resolution in time corresponding to larger spacing between the red points. In practice, one might rather prefer {\it increasing} resolution in time in the final phases of collapse. This might be the case if one imagines $r$ as an input to some more complicated model, for example as an input for a model of galaxy formation (e.g. \citealt{Kitaura_2013}) or a sub-grid model in a hydrodynamic simulation of  gas or chemical evolution in galaxy formation.}
    \label{fig:sampling}
\end{figure}

\section{Discussion}
The \new{above discussion} has shown how one can 
obtain an efficient numerical approximation for the cycloid's evolution \new{using contour integration} coupled with FFTs, \new{just as for} the geometric goat problem \citep{Ullisch}. This \new{is of use} if one \new{requires} the radial evolution of a cycloid \new{(\textit{i.e.} the evolution of a collapsing object)} as a direct numerical lookup table against time. That being said, from the standpoint of numerics, it seems no less efficient to simply generate a grid of $\theta$ and evaluate both $t(\theta)$ and $r(\theta)$ on this grid, and then simply match the elements of these 1-D arrays. 

However, on closer analysis, it is evident that if one wished to have a uniform sampling in time (or for that matter, any arbitrary, user-set sampling), this \new{would not be easily permitted by the latter method}: 
one would need to numerically solve (e.g. using a root-finder) for $\theta$ at every desired $t$ point, and then evaluate $r$ at those $\theta$. This is shown, and further discussed, in Figure \ref{fig:sampling}. In contrast, since the integral solution here, paired with the Fourier method, 
gives $r$ explicitly as a function of $t$, imposing any user-desired sampling in $t$ is trivial.

Finally, we also note that in the context of numerics, our restriction that we can work only on a domain from $\epsilon$ up to $\pi/2 - \epsilon$ in $\theta$ \new{(since we use a contour of radius $\pi/4-\epsilon$)} is not an issue: numerical discretizations always choose sampling points in any case, and the restriction that a sampling point not be in the set of measure zero given by $\{0, \pi/2\}$ is in practice no restriction at all.

\begin{figure}
    \centering
    \includegraphics[width=\linewidth]{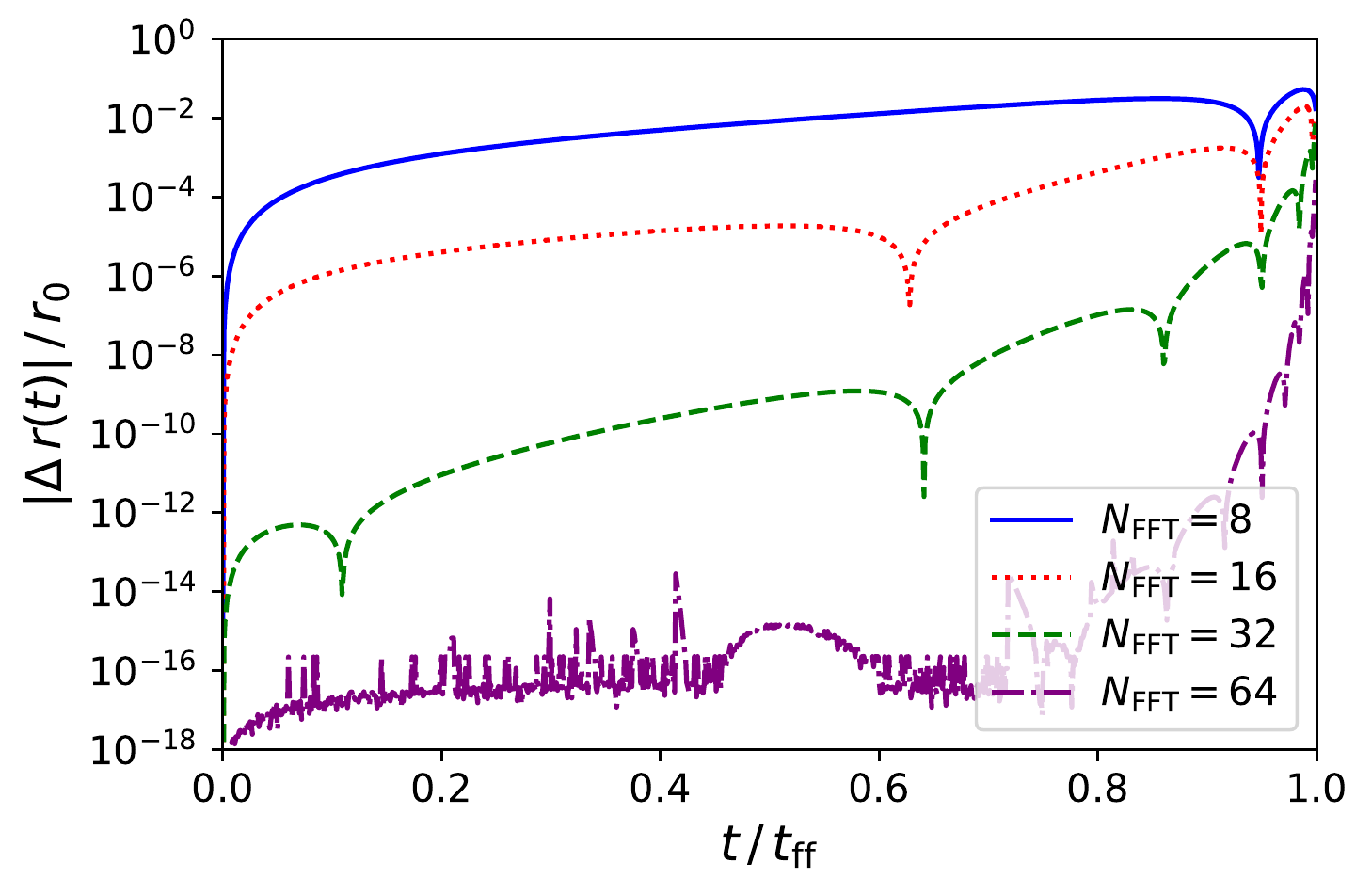}
    \caption{\new{Comparison of root-finding and FFT-based methods to solve the spherical collapse equation. Here, we plot $\left|\Delta r(t)\right|\equiv \left|r^\mathrm{root}(t)-r^\mathrm{FFT}(t)\right|$ against time in dimensionless units, for various choices of the FFT grid-size, $N_\mathrm{FFT}$, assuming $\epsilon = 10^{-4}$. To compute the root-finding solutions, we first solve equation \eqref{eqn:theta_eqn} by numerically finding $\theta$ for a given $t$ (using the Newton-Raphson method) then substituting to find $r(\theta)$. The FFT-based approach offers a speed-up of $\sim$$100\times$ relative to the root-finding, as shown in Figure\,\ref{fig: timings}.}}
    \label{fig: fft-vs-root}
\end{figure}

\begin{figure}
    \centering
    \includegraphics[width=\linewidth]{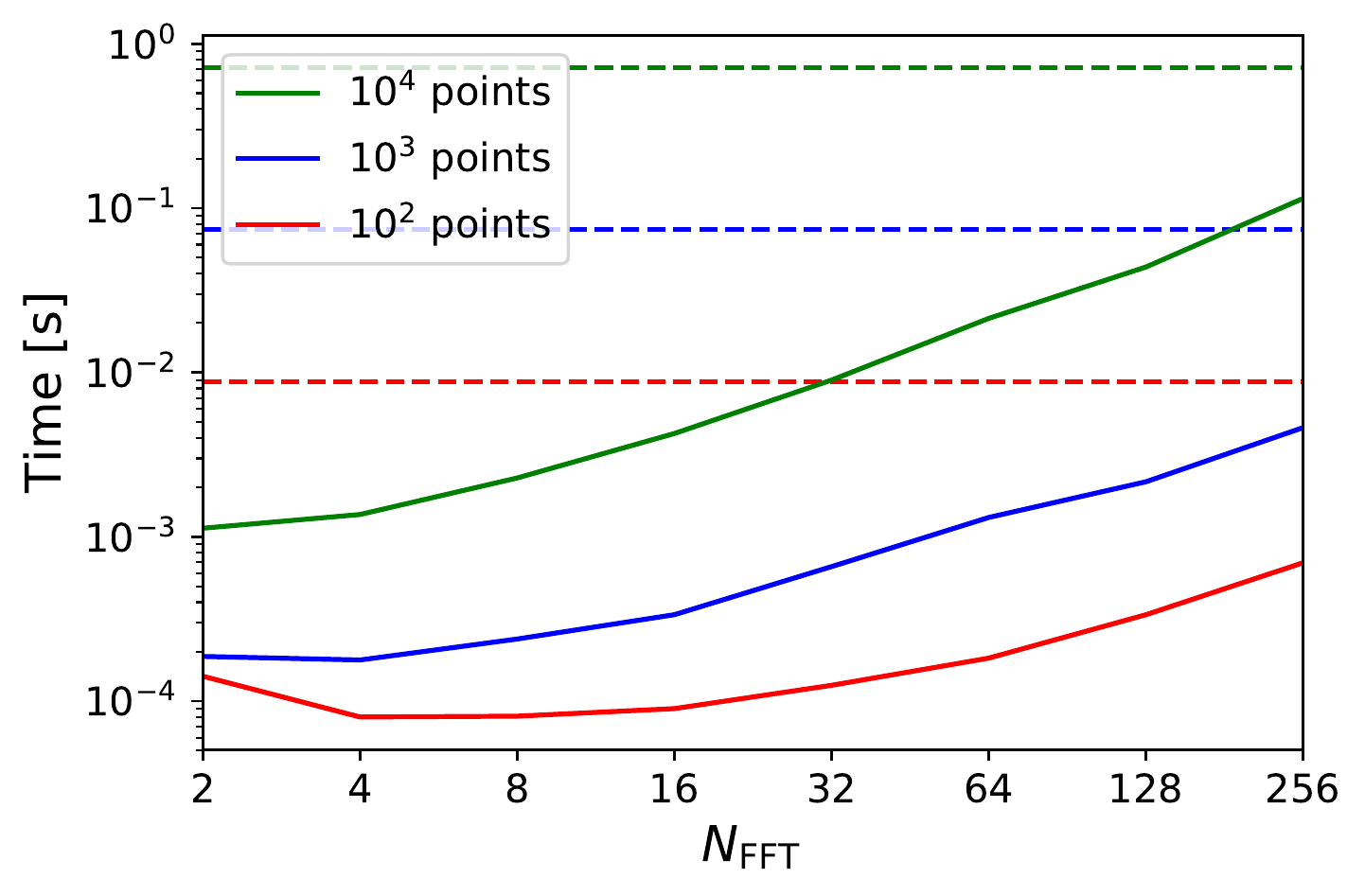}
    \caption{\new{Computation time for the FFT-based spherical collapse solver considered in this work (solid curves) versus the na\"ive root-finding approach discussed in Figure\,\ref{fig: fft-vs-root} (dashed lines). We show results for a range of FFT grid-sizes and three lengths of the input $t$-array, from $10^4$ elements (top line/top curve, green) to $10^2$ elements (bottom line/bottom curve, red). Both methods are simply implemented in \textsc{python}, and the FFT-based approach may be efficiently vectorized.}}
    \label{fig: timings}
\end{figure}

\new{To demonstrate our method, we implement the FFT-based technique in a simple \textsc{python} function.\footnote{This is publicly available on \href{https://gist.github.com/oliverphilcox/559f086f1bf63b23d55c508b2f47bad3}{GitHub}.} Given a set of time co-ordinates $t$ and an array of $2^{N_\mathrm{FFT}}$ sampling points $\{x_i\}$, we compute the coefficients $c_{k}$ \eqref{eqn:ck-def} via a (vectorized) FFT, and thus the corresponding components $r(t)$. Figure \ref{fig: fft-vs-root} compares these to the conventional approach of solving for $\theta$ numerically at each $t$, then using this solution to estimate $r(\theta)$. As $N_\mathrm{FFT}$ increases, our procedure rapidly converges, and we find it to require a $\sim$$100\times$ smaller runtime at $N_\mathrm{FFT}=32$, with only weak dependence on $N_\mathrm{FFT}$ and the size of the time array, as shown in Figure \ref{fig: timings}. Explicitly, the FFT algorithm requires $6\times 10^{-4}\,$s to compute $r(t)$ for $10^4$ points on a single $2.4\,$GHz Intel Skylake CPU, compared with $8\times 10^{-2}\,$s for the root-finding approach. Given the highly optimized FFT libraries that exist, a more mature implementation will be significantly faster.}

\section*{Data Availability}
The data underlying this article will be shared on reasonable request to the corresponding author. A \textsc{python} implementation of our code is available on \href{https://gist.github.com/oliverphilcox/559f086f1bf63b23d55c508b2f47bad3}{GitHub}.

\section*{Acknowledgments}
We thank J.R. Gott, III for initializing ZS's interest in the cycloid problem a number of years ago, and for enlivening the intervening time with sage advice and good humor. OP thanks Jeremy Goodman and William Underwood for insightful discussions regarding complex analysis. OP acknowledges funding from the WFIRST program through NNG26PJ30C and NNN12AA01C. No goats were harmed in the making of this work.

\bibliographystyle{mnras}
\bibliography{bib}

\begin{thebibliography}{}
\makeatletter
\relax
\def\mn@urlcharsother{\let\do\@makeother \do\$\do\&\do\#\do\^\do\_\do\%\do\~}
\def\mn@doi{\begingroup\mn@urlcharsother \@ifnextchar [ {\mn@doi@}
  {\mn@doi@[]}}
\def\mn@doi@[#1]#2{\def\@tempa{#1}\ifx\@tempa\@empty \href
  {http://dx.doi.org/#2} {doi:#2}\else \href {http://dx.doi.org/#2} {#1}\fi
  \endgroup}
\def\mn@eprint#1#2{\mn@eprint@#1:#2::\@nil}
\def\mn@eprint@arXiv#1{\href {http://arxiv.org/abs/#1} {{\tt arXiv:#1}}}
\def\mn@eprint@dblp#1{\href {http://dblp.uni-trier.de/rec/bibtex/#1.xml}
  {dblp:#1}}
\def\mn@eprint@#1:#2:#3:#4\@nil{\def\@tempa {#1}\def\@tempb {#2}\def\@tempc
  {#3}\ifx \@tempc \@empty \let \@tempc \@tempb \let \@tempb \@tempa \fi \ifx
  \@tempb \@empty \def\@tempb {arXiv}\fi \@ifundefined
  {mn@eprint@\@tempb}{\@tempb:\@tempc}{\expandafter \expandafter \csname
  mn@eprint@\@tempb\endcsname \expandafter{\@tempc}}}

\bibitem[\protect\citeauthoryear{{Gunn} \& {Gott}}{{Gunn} \&
  {Gott}}{1972}]{Gunn}
{Gunn} J.~E.,  {Gott} J.~Richard I.,  1972, \mn@doi [\apj] {10.1086/151605},
  \href {https://ui.adsabs.harvard.edu/abs/1972ApJ...176....1G} {176, 1}

\bibitem[\protect\citeauthoryear{{Jackson}}{{Jackson}}{1916}]{Jackson16}
{Jackson} D.,  1916, Ann. of Math., 17, 172

\bibitem[\protect\citeauthoryear{{Jackson}}{{Jackson}}{1917}]{Jackson17}
{Jackson} D.,  1917, Ann. of Math., 19, 142

\bibitem[\protect\citeauthoryear{Kitaura, Yepes  \& Prada}{Kitaura
  et~al.}{2013}]{Kitaura_2013}
Kitaura F.-S.,  Yepes G.,   Prada F.,  2013, \mn@doi [Monthly Notices of the
  Royal Astronomical Society: Letters] {10.1093/mnrasl/slt172}, 439, L21–L25

\bibitem[\protect\citeauthoryear{{Lin}, {Mestel}  \& {Shu}}{{Lin}
  et~al.}{1965}]{LMS65}
{Lin} C.~C.,  {Mestel} L.,   {Shu} F.~H.,  1965, \mn@doi [\apj]
  {10.1086/148428}, \href
  {https://ui.adsabs.harvard.edu/abs/1965ApJ...142.1431L} {142, 1431}

\bibitem[\protect\citeauthoryear{{Luck}, {Zdaniuk}  \& {Cho}}{{Luck}
  et~al.}{2015}]{Luck15}
{Luck} R.,  {Zdaniuk} G.,   {Cho} H.,  2015, International Journal of
  Engineering Mathematics, 523043

\bibitem[\protect\citeauthoryear{Tomita}{Tomita}{1969}]{Tomita}
Tomita K.,  1969, \mn@doi [Progress of Theoretical Physics] {10.1143/PTP.42.9},
  42, 9

\bibitem[\protect\citeauthoryear{Ullisch}{Ullisch}{2020}]{Ullisch}
Ullisch I.,  2020, \mn@doi [The Mathematical Intelligencer]
  {10.1007/s00283-020-09966-}, 42, 12

\makeatother
\end{thebibliography}

\bsp	
\label{lastpage}
\end{document}